\documentclass[a4paper, prd, twocolumn, nofootinbib,superscriptaddress]{revtex4-1}

\usepackage{color}
\usepackage{multirow}
\usepackage{amstext}
\usepackage{amssymb}
\usepackage{amsmath}
\usepackage{hyperref}
\usepackage{paralist}
\usepackage{graphicx}
\usepackage{url}

\def\beq{\begin{equation}}\def\eeq{\end{equation}}
\def\bea{\begin{align}}\def\eea{\end{align}}

\newfont{\cursive}{pzcmi at 9pt}

\begin{document}

\title[]{Inferring the gravitational wave memory for binary coalescence events}

\author{Neev Khera}
\affiliation{Institute for Gravitation and the Cosmos, Pennsylvania State University, University Park, PA 16802, USA}

\author{Badri Krishnan}
\affiliation{Max Planck Institute for Gravitational Physics (Albert Einstein Institute), Callinstrasse 38, D-30167 Hannover, Germany}
\affiliation{Leibniz Universit\"at Hannover, 30167 Hannover, Germany}

\author{Abhay Ashtekar}
\affiliation{Institute for Gravitation and the Cosmos, Pennsylvania State University, University Park, PA 16802, USA}

\author{Tommaso De Lorenzo} 
\affiliation{Institute for Gravitation and the Cosmos, Pennsylvania State University, University Park, PA 16802, USA}



\begin{abstract}

  Full, non-linear general relativity predicts a memory effect for
  gravitational waves. For compact binary coalescence, the total
  gravitational memory serves as an inferred observable, conceptually
  on the same footing as the mass and the spin of the final black
  hole. Given candidate waveforms for any LIGO event, then, one can
  calculate the posterior probability distribution functions for the
  total gravitational memory, and use them to compare and contrast the
  waveforms. In this paper we present these posterior distributions
  for the binary black hole merger events reported in the first
  Gravitational Wave Transient Catalog (GWTC-1), using the
  Phenomenological and Effective-One-Body waveforms.  On the whole,
  the two sets of posterior distributions agree with each other quite
  well though we find larger discrepancies for the $\ell=2, m=1$ mode
  of the memory.  This signals a possible source of systematic errors
  that was not captured by the posterior distributions of other
  inferred observables. Thus, the posterior distributions of various
  angular modes of total memory can serve as diagnostic tools to
  further improve the waveforms.  Analyses such as this would be
  valuable especially for future events as the sensitivity of ground
  based detectors improves, and for LISA which could measure the total
  gravitational memory directly.
  
\end{abstract}

\maketitle

\section{Introduction}
\label{sec:intro}

The detection of gravitational waves enables tests of general
relativity that were not possible using the electromagnetic window.
For example, through observations of compact binary mergers one can
verify
higher order post-Newtonian effects in the inspiral regime, and probe
the nature of the final remnant black hole in the post-merger regime
\cite{PhysRevLett.116.221101,PhysRevLett.123.011102,PhysRevD.100.104036}.
Similarly, in the search for potential deviations from general
relativity, one can use the parameterized post-Newtonian
formalism 
as a general framework in the inspiral regime, and black hole
perturbation theory in the post-merger regime. The merger itself
cannot be addressed by these approximation schemes because non-linear
effects of general relativity are especially important there. On the
other hand, precisely for the same reason, the merger provides a
promising place to look for potential deviations from general
relativity.

So far, there is no generally accepted framework for describing the
merger itself analogous to the parameterized post-Newtonian framework.
Several tests have been proposed in the literature which attempt to
probe different aspects of the merger.  However, to reliably test
whether predictions of general relativity are borne out in
observations, one needs to be confident that the theoretical waveforms
used in these tests capture predictions of the theory to a
sufficiently high degree of accuracy.  Although so far, there is no
generally accepted framework to test the accuracy of theoretical
predictions describing the merger
itself, 
several tests have been proposed in the literature to probe different
aspects of these predictions.  These include, for example, various
consistency checks between the inspiral and the merger
\cite{Ghosh:2017gfp,PhysRevD.97.124069}, and also tests of
phenomenological waveform models for the merger
\cite{PhysRevLett.116.221101}. In this paper we suggest that the total
gravitational memory can be used as a new tool in the same direction.

General relativity predicts that memory associated with gravitational
waves emitted in compact binary coalescences would be generically
non-zero.  For the interferometric gravitational wave detectors, this
corresponds to a permanent displacement of the test masses due to the
flux of 
gravitational waves across the plane of the detector
\cite{Christodoulou:1991cr,Frauendiener_1992,Thorne:1992sdb,Blanchet:1992br,Wiseman:1991ss}.
See also
\cite{Johnson:2018xly,Kennefick:1994nw,Favata:2008ti,Favata:2008yd,Favata:2009ii,Favata:2010zu,Favata:2011qi}
for later work discussing prospects for detecting memory and further
calculations of the memory within the post-Newtonian framework.
Advances in numerical relativity towards calculating memory in
numerical simulations of black hole mergers are given in
\cite{Pollney:2010hs,Mitman:2020pbt}.  A direct measurement of the
memory would be a probe of non-linear aspects of general relativity,
and also of the merger, since the effect is the largest during this
phase.  However, thus far, we do not have a direct measurement either
for single events, or collectively for a population of events
\cite{Hubner:2019sly,Divakarla:2019zjj,Talbot:2018sgr,McNeill:2017uvq,Lasky:2016knh}.
But the total gravitational memory is a bonafide observable in full
general relativity, expressible as a functional of the gravitational
wave strain in a detector.  Therefore, assuming general relativity, it
is possible to \emph{infer} its value.

As we now explain, this inference relies on the waveform model used.
For a binary system, a gravitational waveform received at a detector
is parameterized by at least 10 intrinsic parameters (without
restricting oneself to general relativity): the two component masses
$m_1$ and $m_2$, the individual spins of the two components
$\mathbf{S}_1$ and $\mathbf{S}_2$, and two additional parameters in
case the system is in an eccentric orbit (namely, the eccentricity and
the orientation of the elliptical orbit). The total mass is denoted
$M=m_1+m_2$.  It is conventional to use the dimensionless spin
parameter $\chi_i = \mathbf{S}_i\cdot \mathbf{\widehat{L}}/m_i^2$
($i=1,2$) instead of the spin itself, where $\mathbf{\widehat{L}}$ is
the unit-vector in the direction of the orbital angular momentum
vector $\mathbf{L}$.  Similarly, the effective spin $\chi_{eff}$ which
appears in several waveform models is a weighted sum of the individual
spins: $\chi_{eff} = (m_1\chi_1 + m_2\chi_2)/M$.  The dimensionless
spin components perpendicular to $\mathbf{\widehat{L}}$ are denoted
$\chi_{1,2}^\perp$.  In addition to the masses and spins, we will have
4 extrinsic parameters.  This includes the luminosity distance to the
source $D_L$, and three additional extrinsic parameters which
determine the orientation of the source.  Let us denote the intrinsic
parameters collectively by $\vec{\lambda}$, and the 4 extrinsic
parameters by $\vec{\mathcal{A}}$.  If one or both of the compact
objects is a neutron star, we will have further parameters depending
on the structure of the star.  Given these parameters, we can
determine the gravitational waveforms
$h_{+,\times}(t;\vec{\lambda},\vec{\mathcal{A}})$ for the two
polarizations. Given, in addition, the orientation of the detector
which requires three additional angles collectively denoted
$\vec{\Theta}$, the received strain
$h(t;\vec{\lambda},\vec{\mathcal{A}},\vec{\Theta})$ in a detector can
be calculated by a suitable projection of $h_{+,\times}$.  One can
show that one of the angles in $\vec{A}$ is degenerate with an angle
in $\vec{\Theta}$, namely, the polarization angle (see
e.g. \cite{Harry:2010fr}).  Consequently, there are only three
independent parameters in $\vec{\mathcal{A}}$: the luminosity distance
$D_L$, the inclination angle $\iota$ which is the angle between the
source axis and the line of sight to the detector, and a phase
$\varphi_0$.  The three parameters in
$\vec{\Theta}$ are the sky-location of the source in the detector
frame $(\theta,\phi)$, and the polarization angle $\psi$.

Given the measurement of the strain in a detector, one can match
the most accurate available model and determine the values or, more
precisely, the posterior probability distributions, of the intrinsic
parameters $\vec{\lambda}$. These distributions are in fact among the
most important results of the observation, providing us with the measured 
values of these parameters that describe the binary. Note that the above
discussion assumes that the wavelength of the signal is much longer
than the length of the detector arms, and that the signal duration in
much shorter than a sidereal day.  Thus the distinction between
intrinsic and extrinsic parameters is more complicated for e.g. the
LISA detector, but we shall not consider this detail here.

Once we have the probability distribution for $\vec{\lambda}$,
assuming general relativity, we can use it to calculate values of
other important observables associated with the binary. Following a
general convention, we will refer to the values (or rather, the
probability distributions) of $\vec{\lambda}$ as \emph{measured}
quantities, and those of additional observables that can then be
deduced as \emph{inferred} values.  The most widely used inferred
observables for a binary are the mass $M_f$, the spin $\mathbf{S}_f$
and the recoil (or kick) velocity $\mathbf{v}$ of the remnant.  The
total gravitational wave memory is on a similar footing as these:
given the waveform parameters and a particular waveform model, values
of various modes in the angular decomposition of the memory can be
uniquely inferred assuming general relativity.

The first goal of this paper is to carry out this procedure in detail
and to obtain the posterior distributions of the memory modes for the
binary black hole merger events reported in the first Gravitational
Wave Transient Catalog (GWTC-1) \cite{LIGOScientific:2018mvr}.  Now,
the properties of the commonly used inferred observables -- the final
black hole parameters such as $M_f,\mathbf{S}_f, \mathbf{v}$-- have
important astrophysical and theoretical applications.  Gravitational
memory is likely not of direct astrophysical interest. Nonetheless,
since it is a genuine observable in general relativity, it has
interesting theoretical implications. In particular,
\emph{differences} in the memory for different waveform models are
significant.  If for example, for a given event, the memory turns out
to be statistically different for different waveform models, they
cannot both be accurate approximations to exact general
relativity. Therefore the statistical difference would point to a
difference between the underlying \emph{physical} assumptions of the
models, indicating that these models can be further improved. As
detectors become increasingly sensitive, these differences might
become more significant and can play a useful role in improving
waveform models.  The second goal of this paper, then, is to advocate
the use of the memory as a diagnostic tool for investigating physical
differences between different waveform models.

The plan for the rest of the paper is the following.  In
Sec.~\ref{sec:constraints} we shall explain the basic formalism for
calculating the linear and non-linear parts of the memory.
Sec.~\ref{sec:memory} applies this to the events published by the LIGO
and Virgo collaborations and finally Sec.~\ref{sec:discussion}
concludes with a discussion of the results and possible future
applications of the memory.

\section{Constraints on gravitational waveforms and the memory}
\label{sec:constraints}

We begin with a description of the gravitational wave signal emitted
by a compact binary source.  
The starting point for understanding the behavior of gravitational
radiation in numerical relativity and gravitational waveform modeling
is the Weyl tensor component
$\Psi_4 = C_{abcd}n^a\bar{m}^bn^c\bar{m}^d$.  Here $C_{abcd}$ is the
Weyl tensor, and $(l^a,n^a,m^a,\bar{m}^2)$ is a suitably chosen
null-tetrad adapted to spheres centered on the source.  Thus, $l^a$
and $n^a$ are respectively the outgoing and ingoing null normals to
these spheres, while the complex vector field $m^{a}$ is tangential to
the spheres and adapted to the source axis, and $\bar{m}^a$ is the
complex conjugate of $m^a$.%
\footnote{For a non-precessing system, the source axis would be the
  direction of the orbital angular momentum, while for a precessing
  system the direction of the total angular momentum provides an
  approximately conserved direction.}
%
The only non-vanishing inner-products are $m\cdot\bar{m} = 1$ and
$l\cdot n = -1$.  The notion of spin-weight plays an important
role.  This refers to the behavior of quantities under `spin
rotations' $m\rightarrow e^{i\psi}m$.  A quantity $F$ is said to have
spin-weight $s$ if $F\rightarrow e^{is\psi}F$ under this
transformation.  Thus, $m^a$ itself has spin weight $+1$ while
$\bar{m}^a$ has $s=-1$.  The Weyl tensor component $\Psi_4$ has
$s=-2$.

The two polarizations of the gravitational wave strain $h_{+,\times}$ are related to $\Psi_4$ according to
\begin{equation}
  \Psi_4 = -\ddot{\mathfrak{h}} \quad \textrm{where} \quad \mathfrak{h} := h_+-ih_\times\,.
\end{equation}
The emitted gravitational wave signal at large
distances from the source can be expanded in terms of spin-weighted
spherical harmonics:
\begin{equation}
  \label{eq:hmode}
  \mathfrak{h} = \frac{1}{D_L}\,\,\sum_{\ell=2}^\infty\sum_{m=-\ell}^{\ell} 
  \mathfrak{h}_{\ell m}(t;\vec{\lambda})\,\, {}_{-2}Y_{\ell m}(\iota,\varphi_0)\,.
\end{equation}
Here ${}_{-2}Y_{\ell m}$ is a spin-weighted spherical harmonic of spin
weight $-2$, and $D_L$ is the luminosity distance from the source.
See e.g. \cite{Reisswig:2010di,Berti:2007fi} for a discussion of the integrations in
time required to go from $\Psi_4$ to $\dot{\mathfrak{h}}$ and eventually to
$h_{+,\times}$.

As explained in Section \ref{sec:intro} , $h_{+,\times}$ are functions
of time, and they are parameterized by
$(\vec{\lambda},\iota,\varphi_0,D_L)$.  The signal $h(t)$ seen at a
detector is
\begin{eqnarray}
  h(t;\vec{\lambda},\vec{A},\Theta) &=& F_+(\theta,\phi,\psi)h_+(t;\vec{\lambda},\vec{A}) \nonumber \\
                                    &+& F_\times(\theta,\phi,\psi)h_\times(t;\vec{\lambda},\vec{A})\,.
\end{eqnarray}
Here $F_{+,\times}$ are the detector beam pattern functions.

Numerical simulations provide us with the mode amplitudes $h_{\ell m}$
for a selected set of points in parameter space and for a few chosen
modes for which $\Psi_4$ can be extracted
reliably. 
For analyzing gravitational wave signals, it is much more practical to
construct analytical models that interpolate between these chosen
points in parameter space, and 
then use 
these models for gravitational wave mode amplitudes.  Significant
advances have been made in addressing this interpolation problem (see
e.g. \cite{Setyawati:2019xzw} for recent work in this direction).  Two
particular waveform models have been used extensively for interpreting
gravitational wave data.  The first is the Effective-One-Body (EOB)
framework originally suggested in \cite{Buonanno:1998gg}; see
\cite{Damour:2016bks} for a review and
e.g. \cite{Ossokine:2020kjp,Bohe:2016gbl,Taracchini:2013rva,Pan:2013rra,Pan:2011gk,Nagar:2018zoe,Damour:2014sva}
for further developments.  The second commonly used models are the
so-called Phenomenological models originally proposed in
\cite{Ajith:2007kx}; see
e.g. \cite{Khan:2019kot,Khan:2018fmp,London:2017bcn,Khan:2015jqa,Husa:2015iqa,Santamaria:2010yb}
for further developments.  It is beyond the scope of this article to
review the basic ideas underlying these models, comparisons between
them, and their relative strengths and weaknesses.  Rather, we will
use them to perform `null tests' by first assuming that they both
correctly capture general relativity, with sufficient accuracy for
detection and parameter estimation in binary mergers, and then
comparing their predictions for other observables as a diagnostic tool
for potential systematic errors.

Our analysis is based on an infinite tower of constraints on gravitational
waveforms, imposed by certain `balance laws' in full, non-linear general 
relativity. Let us begin with the easier cases, namely the balance laws for 
energy  $E$ and linear momentum $P_{i}$. Let us first note that the total 
fluxes, $\Delta E$ and $\Delta P_i$, carried away by the gravitational 
waves are given by
%
%
\begin{eqnarray}
  \Delta E &=& \frac{D_L^2c^3}{16\pi G}\int_{-\infty}^{\infty} \!\!\!dt \,\oint d\Omega\, \,
  |\dot{\mathfrak{h}}|^2\, ,  \label{eq:energy_flux}\\
  \Delta{P}_i &=& \frac{D_L^2c^2}{16\pi G}\int_{-\infty}^{\infty} \!\!\!dt\, \oint d\Omega\,\, \hat{x}_i(\iota,\varphi_0)| \,\dot{\mathfrak{h}}|^2 \,\, .  \label{eq:momentum_flux}
\end{eqnarray}
Here $\hat{x}_1 = \sin\iota \cos\varphi_0$,
$\hat{x}_2 = \sin\iota \sin\varphi_0$, $\hat{x}_3 = \cos\iota$ and
$d\Omega = \sin\iota\,d\iota\,d\varphi_0$. %
\footnote{A corresponding formula also exists for the flux of angular momentum, but it involves several subtle issues \cite{Ashtekar:2019rpv}; we shall not discuss it in this article.} 
Note that the fluxes $\Delta E$ and $\Delta P_{i}$ are completely determined by the waveform $\mathfrak{h}$. Therefore, given the initial (i.e., ADM) mass and the waveform, the balance laws determine the energy-momentum of the final black hole from which one can extract its mass $M_{f}$ and its recoil velocity $\mathbf{v}$. 

Now, since the radiated energy, the recoil velocity, and the final
mass are all parameters of direct astrophysical interest, there is an
extensive literature on calculating these quantities as functions of
the initial parameters
\cite{Healy:2016lce,PhysRevLett.122.011101,Zappa:2017xba}.  These
functions are typically obtained as fits to the results of numerical
simulations. But as indicated above, we can also calculate these
quantities using the model waveforms and the initial parameters that
label them. If these waveforms are to accurately represent general
relativity, the answers must agree with the fits from numerical
relativity.  Note that it is not obvious that the two calculations
must necessarily agree.  As an example of the gap between the two
calculations, consider the mass $M_{f}$ and the spin $\mathbf{S}_{f}$
of the final black hole.  In numerical relativity, these are typically
calculated using geometrical fields on black hole \emph{horizons}
rather than waveforms in the \emph{asymptotic regions} (see
e.g. \cite{Dreyer:2002mx,Owen:2009sb}).  While one expects the two
sets of values to agree at late times, their equality has not yet been
established mathematically (because of technical issues concerning the
structure at future timelike infinity $i^{+}$).  Therefore, a
comparison between the two would serve as a useful check on overall
consistency.  In addition, as we discuss in Section \ref{sec:memoryA},
one can view such comparisons as accuracy tests for the waveforms.
Any disagreements, even if not significant for current gravitational
wave data analysis purposes, might point directions leading to
improved waveform models.  Eventually, as detectors improve in
sensitivity, accuracy requirements on the waveforms become more
stringent.  Thus, such improvements might be part of the various
ingredients in waveform modeling necessary in the coming era of high
precision gravitational wave astronomy.

The main focus of this paper is on the non-trivial constraints on
the waveforms obtained from the fluxes of \emph{supermomenta} which, as we
shall now see, are closely connected with the total gravitational memory,  which is given by
\begin{equation}
  \Delta \mathfrak{h} (\iota,\varphi_0) = \left.\mathfrak{h}\right|_{u=\infty} - \left.\mathfrak{h}\right|_{u=-\infty}\, .
\end{equation}
In practice, $\mathfrak{h}$ can be calculated by performing two
time-integrals of $\Psi_{4}$.  This procedure involves two integration
constants \cite{Reisswig:2010di,Berti:2007fi}. The first vanishes in
binary coalescences, since the Bondi news $\dot{\mathfrak{h}}$ goes to
zero in the distant past as well as distant future (just as $\Psi_{4}$
does).  The second is generally used to set
$\mathfrak{h}|_{u=-\infty}\,=0$.  However, since the total memory
$\Delta \mathfrak{h}$ is a \emph{difference}, its value is independent
of the choice of this integration constant; it is a well-defined
observable in general relativity, without any further inputs.

The value of $\Delta \mathfrak{h}$ is governed by the supermomentum
balance laws, associated with supertranslations. More precisely,
presence of gravitational waves in the asymptotic region forces one to
enlarge the 4 dimensional group of translations of flat space-time to
an infinite dimensional group of ``angle dependent translations'',
called \emph{supertranslations} \cite{Bondi:1962px,Sachs:1962wk}.
Just as there are energy momentum balance laws associated with
asymptotic translations, Einstein's equations imply that there are
supermomentum balance laws associated with supertranslations
\cite{Ashtekar:1981bq}.  As shown in \cite{Ashtekar:2019viz}, under
assumptions that are normally made in the analysis of compact binary
coalescence, they imply:
\begin{eqnarray}
  \eth^2\Delta \mathfrak{h} &=& - \frac{2G}{D_L c^2}\Big( M - \frac{M_f}{\gamma^3(1- \mathbf{v}\cdot\hat{x}/c)^3}\Big) \nonumber \\
                            &+& \frac{D_L}{2 c}\int_{-\infty}^\infty\!\!\! dt |\dot{\mathfrak{h}}|^2\! \, .
                                \label{eq:constraint}
\end{eqnarray}
Here, $M$ is the total initial mass of the system; $M_f$, the mass of the final black hole; $\mathbf{v}$, the recoil velocity; 
and, $\eth$, the angular derivative, whose action on a scalar $F$ with spin weight  $s$ is a spin weight $s+1$ scalar, given by \cite{GMS,Goldberg:1966uu}
\begin{equation} \label{eq:eth}
  \eth F := -\frac{1}{\sqrt{2}}(\sin\iota)^s\left(\frac{\partial}{\partial\iota} + \frac{i}{\sin\iota}\frac{\partial}{\partial\varphi_0} \right)\left(\frac{F}{(\sin\iota)^s} \right) \, .
\end{equation}
Since $\Delta \mathfrak{h}$ is the strain, it has spin weight
$-2$. Thus, the left-hand-side of Eq.~(\ref{eq:constraint}) has
spin-weight 0, consistent with the right-hand-side.

Note that both sides of Eq.~(\ref{eq:constraint}) have a
$(\iota, \varphi_0)$ dependence. Therefore, we can carry our a mode
decomposition of this equation using spherical harmonics
\begin{eqnarray}  \label{eq:ellm}
C_{\ell} (\Delta \mathfrak{h})_{\ell, m} &=& - \frac{2G}{D_{L}c^{2}} \Big(M\, -
\frac{M_{f}}{\gamma^{3}(1- \frac{\mathbf{v}}{c}\cdot \hat{x})^{3}}\Big)_{\ell,m} \nonumber \\ 
&+&\, \frac{D_L}{2 c}\, \Big(\int_{-\infty}^\infty\!\!\! dt |\dot{\mathfrak{h}}|^2\Big)_{\ell,m} 
\end{eqnarray}
where 
\begin{equation} \label{Cell}
C_{\ell} = \frac{1}{2} \, (\ell-1) \ell (\ell+1) (\ell+2)\, .
\end{equation}

The $\ell=0$ and $\ell=1$ components of (\ref{eq:ellm}) provide us the
balance laws for energy and momentum.  (Note that $C_{\ell} =0$ in
these cases.)
Now,  as we remarked above, if we know the initial mass $M$, expressions (\ref{eq:energy_flux}) and (\ref{eq:momentum_flux}) of the energy and momentum flux, together with the balance laws 
(\ref{eq:ellm}) for $\ell=0,1$ can be used to determine the mass $M_f$ and the recoil velocity 
of the final black hole.   Once $M, M_{f}$ and $\mathbf{v}$ are known, the only other field in Eq.~(\ref{eq:ellm})  is the waveform $\mathfrak{h}$. Therefore, the $\ell \ge 2$ modes of Eq.~\ref{eq:ellm} provides an infinite tower of constraints to be tested on the $\mathfrak{h}$ provided by waveform models.   

However, currently the models do not incorporate the total memory $\Delta \mathfrak{h}$ that appears on the left side of (\ref{eq:ellm}), whence the constraints are violated for $\ell \ge 2$.  But the right hand side is dominated by aspects of the waveform that, one expects,  are well modeled.  For example the leading contribution to the right hand side comes from the $(2,\pm 2)$ mode of the waveform, which all waveform models incorporate. Hence we can turn around the constraint, and \emph{use it} to calculate the memory using the well modeled aspects of the waveform. Thus, Eq.~(\ref{eq:ellm})  serves as the primary equation which determines the total memory and its mode decomposition.  (The first term on the right side is often called the ``linear'' memory and the second term the ``non-linear'' memory.)

\section{Inferred memory for the observed events}
\label{sec:memory}

For the observed events, parameter estimation using EOB and Phenom models provides us with two posterior distributions for the initial parameters. Given either of them and the corresponding waveform, we can calculate the probability distribution of different modes of the memory.  We can do this by taking sample points from the posterior, and then using Eq.~(\ref{eq:ellm}) to calculate the memory for each sample. This would then give us two posterior distributions of the inferred memory for any given event. 
We can then check if the the differences in the inferred memory
between the two models is less than the statistical errors. In all of
the following we use the IMRPhenomPv2 and SEOBNRv3 for events
published in the GWTC-1
catalog.

The first step in this procedure is to use the energy-momentum flux to calculate the remnant parameters. This procedure provides us with a posterior probability distribution for the mass and the recoil velocity of the final black hole.  We carry out this step in Section \ref{sec:memoryA}.
In Section \ref{sec:memoryB}, we use these values in conjunction with
Eq. (\ref{eq:ellm}) to arrive at a streamlined procedure to calculate
the inferred values of various angular modes of the memory.  In
Section \ref{sec:memoryC} this procedure is used to obtain the
probability distributions for the leading memory modes in the GWTC-1
events.  Differences between these distributions can be used as
diagnostic tools to detect potential discrepancies and further improve
the waveforms.

\subsection{The remnant mass $M_{f}$ and recoil velocity $\mathbf{v}$}
\label{sec:memoryA}

As we discussed in Section \ref{sec:constraints}, given a waveform
model we can use Eq.~(\ref{eq:energy_flux}) and
(\ref{eq:momentum_flux}) to calculate the remnant mass $M_f$ and
recoil velocity $\mathbf{v}$.  One can then compare these values with
the fits to masses and recoil velocities provided by numerical
relativity using fields at the horizon \cite{PhysRevLett.122.011101}
and the posterior probability for the input parameters provided by the
model. This comparison provides a first check on the model
waveforms. Assuming that each waveform agrees with the corresponding
numerical relativity prediction, one can compare the predictions of
the two waveform models.  As we will see, this comparison can bring
out the differences between the models, thereby providing guidance for
further improvements.  Determination of $M_f$ and $\mathbf{v}$ will
also serve a second purpose: Knowing their values, we will be able to
calculate the $\ell \ge 2$ components of the memory in Section
\ref{sec:memoryB}.

\begin{figure}
  \centering    
  \includegraphics[width=\linewidth]{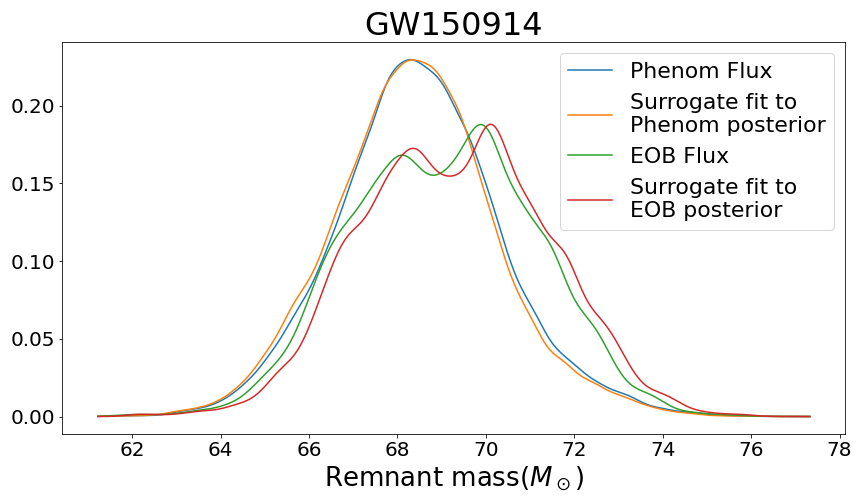}
  \caption{GW150914: distribution of the final mass using the
    IMRPhenomPv2 and SEOBNRv3 models, compared with the numerical
    relativity fits applied to each of the posterior distributions.
    Unlike the kick-velocity results shown in
    Fig.~\ref{fig:gw150914_remnant_kick}, the results for the final
    mass using the two models is mostly consistent with the numerical
    relativity results. The reason is that the energy flux is
    dominated by the $\ell=2$ modes which are accurately modeled.}
  \label{fig:gw150914_remnant}
\end{figure}
\begin{figure}
  \centering    
  \includegraphics[width=\linewidth]{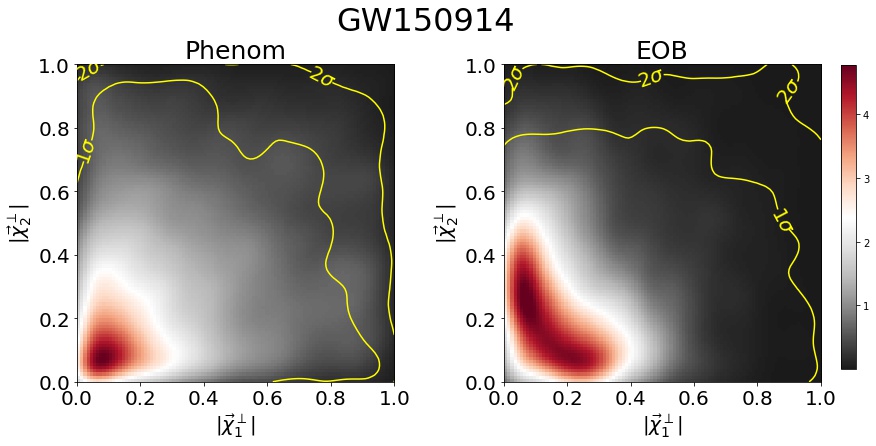}
  \caption{GW150914: Posterior distributions for the perpendicular
    dimensionless spin components $\chi^\perp_{1,2}$. The Phenom model
    uses a single effective spin while the EOB is parameterized by the
    individual spins.  This gives rise to the bimodal distribution for
    $\chi_{1,2}^\perp$ in the EOB model, which in turn leads to the
    double hump in the posterior distribution of the final mass in
    Fig.~\ref{fig:gw150914_remnant}.}
  \label{fig:gw150914_remnant_spins}
\end{figure}
\begin{figure*}
  \centering    
  \includegraphics[width=0.9\linewidth]{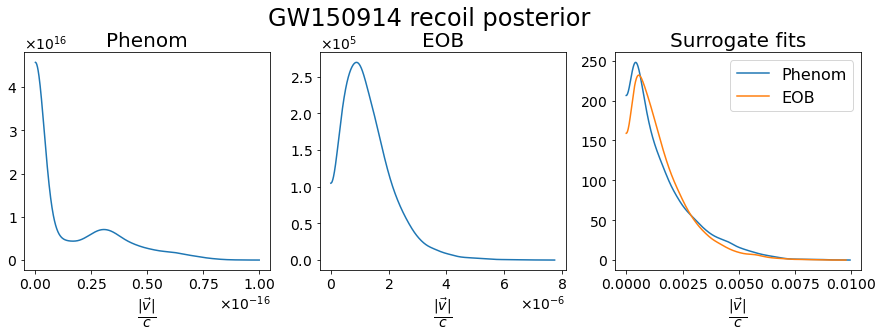}
  \caption{GW150914: Posterior distribution for the recoil velocity
    (in units with $c=1$) using the IMRPhenomPv2 (left panel) and
    SEOBNRv3 (middle panel) models, and also accurate fits to
    numerical relativity calculations (right panel).  We note large
    discrepancies between both the models and the accurate numerical
    relativity results; the distribution in the left panel has the
    bulk of its support for $v/c \sim \mathcal{O}(10^{-16})$ (making
    it consistent with being purely numerical noise), the middle panel
    for $\mathcal{O}(10^{-6})$, while the correct answer in the right
    panel has $\mathcal{O}(10^{-3})$.  As discussed in the text, this
    is not surprising because neither of the models attempt to model
    the higher modes necessary for the kick velocity. However, the
    discrepancy between the Phenom and EOB models is also interesting
    to note.}
  \label{fig:gw150914_remnant_kick}
\end{figure*}

To determine $M_f$ and $\mathbf{v}$, we need to use the flux
expressions Eq.~(\ref{eq:energy_flux}) and
(\ref{eq:momentum_flux}). However, there is a subtlety: Since
waveforms are readily available only for finite time intervals, in
practice we have to truncate the time integrals in flux expressions to
finite intervals.
In the distant future, truncation can be carried out readily without
incurring excessive errors because the waveform decays exponentially.
However, in the past the flux falls-off slowly. For the Phenom model
the waveform is available for sufficiently early times and hence the
error due to truncation can be made negligible. However for the EOB
model, we are unable to generate the waveform at sufficiently early
times; the implementation in the LAL Simulation software library
\cite{Veitch:2014wba, lalsuite} currently requires the reference frequency and
the starting frequency to be identical. The reference frequency --the
frequency at which time dependent parameters are quoted-- used in the
GWTC-1 catalog is 20Hz.  It is somewhat more complicated to go to
lower frequencies, and we will leave this to future work.  While the
lower frequencies are unimportant for detection and parameter
estimation, their contribution to \emph{total} radiated energy
(\ref{eq:energy_flux}) is not negligible. Therefore, to reduce this
truncation error we will add the energy radiated away from 0Hz to 20Hz
to 0PN order. This can be calculated analytically using the formula
\begin{equation}
\Delta E_{{\rm 0PN}} = M\,\frac{\nu}{2} \,(\pi G M f_{\rm start})^{2/3}
\label{eq:E_0PN}
\end{equation}  
which describes the 0PN radiated energy $\Delta E_{{\rm 0PN}}$ from
retarded time $-\infty$ to when the system reaches a frequency of
$f_{\rm start}$. Here $\nu=m_1m_2/M^2$ is the symmetric mass ratio.
\footnote{For GW150914, $\Delta E_{\rm 0PN}$ radiated till $20\,$Hz,
  gives a surprisingly large value –about $\sim 20\%$ of the total
  radiated energy– because this phase encompasses a large number of
  cycles during which the waveform amplitude is not negligible. By
  contrast, $\Delta E_{\rm 0PN}$ radiated till $1\,$Hz –the cutoff
  frequency used for Phenom models is only $3\%$ of the total
  radiated energy.}  Using this procedure, for each of the two models
we can calculate the remnant parameters in two different ways: (i)
using the energy and momentum fluxes for each waveform model, and,
(ii) using the numerical relativity fits on the posterior distribution
of the input parameters, as determined from the respective
model. Adding (\ref{eq:E_0PN}) significantly improves the agreement between final mass calculated from EOB flux and the fits.

This procedure was carried out for all events considered in this
paper.  Fig.~\ref{fig:gw150914_remnant} illustrates the results with
the posterior distribution of $M_{f}$ in the case of GW150914.  This
is based on the posterior samples available in the GWTC-1 catalog
\cite{LIGOScientific:2018mvr}, for both the IMRPhenomPv2 and SEOBNRV3
waveform models; these are all described in further detail in
Sec.~\ref{sec:memoryC}.

For each model, there is an excellent match between (i) the final mass
as calculated from the flux and (ii) the NR fits from same
posteriors. 
However there is a slight disagreement between the two
models. Although it is not statistically significant, there is an
interesting qualitative difference: While the Phenom plots are a
near-Gaussian, the EOB plots show a `double hump'.
The origin of this difference lies in the differences in parameter
estimation of the 2 models, particularly in the way precession is
treated. This can be seen in Fig.~\ref{fig:gw150914_remnant_spins}
which shows the posterior distributions of the two components of the
dimensionless spins, perpendicular to the orbital angular
momentum. The source of the bimodality of the EOB final mass can be
traced back to that in this posterior distribution which also shows
two modes. Taking samples from each mode in
Fig.~\ref{fig:gw150914_remnant_spins} and comparing with the
corresponding points for the EOB distribution in
Fig.~\ref{fig:gw150914_remnant} reveals that the peaks in each of
these distributions are correlated.  The bottom right peak in
Fig.~\ref{fig:gw150914_remnant_spins} corresponds to the higher peak
in Fig.~\ref{fig:gw150914_remnant}, while the top left peak in
Fig.~\ref{fig:gw150914_remnant_spins} correspnds to the lower peak in
Fig.~\ref{fig:gw150914_remnant}.  In the Phenom model on the other
hand, the double hump is absent both in the posterior of spin
distributions and the posterior distribution of the inferred
$M_{f}$. These differences are all within $1\sigma$ of the
distributions, and therefore they are not statistically
significant. However this points to differences that might become
significant with even louder events that we are likely to see as the
sensitivity of the detectors increases.

The recoil velocity $\mathbf{v}$, on the other hand, shows completely different behavior in each model depending on whether it is calculated using  (i) the momentum flux, or (ii) using numerical fits.  
In addition, the values predicted in the two models using momentum flux are also quite different as seen in Fig.~\ref{fig:gw150914_remnant_kick}. While the kicks from the flux of Phenom are below numerical precision,  in EOB the norm of the kick is: $v/c\, \sim10^{-6}$. The NR fits, by contrast, yield a \emph{much larger} kick, $v/c\,~10^{-3}$. However, the disagreement between the models and NR is not surprising because neither model contains the `higher modes' that are important for calculating the kick.  Fortunately, for calculation of memory $(\Delta \mathfrak{h})_{\ell m}$ in Section \ref{sec:memoryB}, this discrepancy does not play a role because even with a recoil velocity $v/c\, \sim 10^{-3}$, the first term on the right hand side of (\ref{eq:ellm}) that contains the recoil velocity is negligible compared to the second term. For definiteness, 
we will present all results using the recoil velocity as calculated by the fluxes of the waveform model being used.

\subsection{Memory}
\label{sec:memoryB}

For an elliptically polarized gravitational wave one can choose a
frame (aligned with the principal polarization axes) in the plane
transverse to the direction of propagation such that $h_{+,\times}$
are given by
\begin{eqnarray}
  h_+ &=& \eta(t)\left( \frac{1+\cos^2\iota}{2} \right)\cos(2\varphi_0 + 2\varphi(t)) \,, \\
  h_+ &=& \eta(t)\cos\iota\sin(2\varphi_0 + 2\varphi(t)) \,.
\end{eqnarray}
Here $\eta(t)$ is a slowly varying amplitude, $\varphi(t)$ is the
orbital phase, and $\varphi_0$ is an initial phase.  Before we apply
the described procedure to get the distribution of memory generally,
consider the simple case of an absence of precession and higher
modes. The waveform is dominated by the $\ell=2, m=\pm 2$ modes.  In
this case, the complex combination
$\mathfrak{h} = h_{+} -i h_{\times}$ is given by a combination of the
$(2,\pm 2)$ spin-weighted spherical harmonics:
\begin{eqnarray}
  \mathfrak{h} &=& e^{2i\varphi_0}\frac{(1+\chi)^2}{4}\mathfrak{h}_0(t) + e^{-2i\varphi_0}\frac{(1-\chi)^2}{4}\mathfrak{h}^\star_0(t) \nonumber \\
               &\propto& {}_{-2}Y_{2,2}(\iota,\varphi_0)\mathfrak{h}_0(t) + {}_{-2}Y_{2,-2}(\iota,\varphi_0)\mathfrak{h}^\star_0(t)\,.
\end{eqnarray}
where $\mathfrak{h}_0 = \eta(t)e^{-2i\varphi(t)}$ and
$\chi:= \cos\iota$.
When we calculate $|\dot{\mathfrak{h}}|^2$ in the right hand side of
Eq.~(\ref{eq:ellm}), we will obtain products of ${}_2Y_{2,\pm2}$ which
can be expanded in terms of the standard spherical harmonics including
in particular $Y_{20}(\iota,\varphi_0)$ and $Y_{40}(\iota,\varphi_0)$.
Since the original waveform model does not include these modes, the
model waveform does not satisfy Eq.~(\ref{eq:ellm}) and therefore it
is not consistent with general relativity.  To address this situation,
an obvious approach is to modify the waveform model by adding these
specific memory modes. This process can be continued iteratively and
will converge.  In practice, as we shall shortly discuss, the first
iteration will suffice and we shall not consider higher iterations in
this paper.

The same procedure works for a more general model consisting of higher
modes. In general, given that $|\dot{\mathfrak{h}}|^2$ has spin weight
0, it can be expanded as
\begin{equation}
  \label{eq:hdotsq}
  |\dot{\mathfrak{h}}|^2 = \sum_{\ell=0}^\infty\sum_{m=-\ell}^\ell \alpha_{\ell m}Y_{\ell m}(\iota,\varphi_0)\,.
\end{equation}
The coefficients $\alpha_{\ell m}$ appearing in this expansion can be
written in terms of the $3j$-symbols as
\begin{widetext}
  \begin{eqnarray} \label{eq:alphalm}
    \alpha_{\ell m} &=&  \sum_{\ell_1,\ell_2 = 2}^\infty\sum_{m_1=-\ell_1}^{\ell_1}\sum_{m_2 = -\ell_2}^{\ell_2} \mathfrak{h}_{\ell_1,m_1}\mathfrak{h}_{\ell_2,m_2}^\star \oint  {}_{-2}Y_{\ell_1 m_1}(\iota,\varphi_0) {}_{-2}Y_{\ell_2 m_2}^{\star}(\iota,\varphi_0) Y_{\ell m}^\star(\iota,\varphi_0)\, d\Omega \nonumber \\
                    &=&  \sum_{\ell_1,\ell_2 = 2}^\infty\sum_{m_1=-\ell_1}^{\ell_1}\sum_{m_2 = -\ell_2}^{\ell_2} \mathfrak{h}_{\ell_1,m_1}\mathfrak{h}_{\ell_2,m_2}^\star (-1)^{m+m_2}\sqrt{\frac{(2\ell_1+1)(2\ell_2+1)(2\ell+1)}{4\pi}}
                        \begin{pmatrix}

                          \ell_1 & \ell_2 & \ell \\
                          m_1 & -m_2 & m
                        \end{pmatrix}
                        \begin{pmatrix}

                          \ell_1 & \ell_2 & \ell \\
                          -2 & 2 & 0
                        \end{pmatrix}\,.
  \end{eqnarray}
\end{widetext}
With the right hand side of Eq.~(\ref{eq:ellm}) now understood, it is
straightforward to finally obtain the memory
$\Delta\mathfrak{h}$. This can be projected onto a particular detector
response function, though we shall not do so here. 

Given any waveform model, we have thus a straightforward procedure to
calculate the final mass and recoil velocity, as well as the memory.
An important point is that all known waveform models are incomplete in
two respects: (i) the waveform is truncated in practice to fi- nite
time/frequency intervals; and, (ii) not all modes are included in the
model.

The truncation to finite time/frequency intervals throws out the early
inspiral region of the waveform. We saw in Sec.~\ref{sec:memoryA} that
this can lead to significant errors because radiated energy
converges rather slowly in the past. Similarly, certain modes of the
memory converge slowly and are thus prone to significant truncation
errors. To reduce these errors, we employ the same strategy as in
Sec.~\ref{sec:memoryA}: we add the leading order post-Newtonian
contribution to the low frequency portion of the integral,
analytically. More precisely, in the expression of the dominant
contributions to memory given in \cite{garfinkle2016simple}, we
substitute the right hand side of Eq.~(\ref{eq:E_0PN}) for the
radiated energy to obtain the contribution from 0Hz to the starting
frequency $f_{\rm start}$. Finally, by comparing the memory calculated
with varying points of truncation, we estimate the corresponding
error. All results reported have a starting frequency of 20Hz for EOB
and 1Hz for Phenom, and errors much smaller than the standard de-
viation. The reason for the truncating EOB is at 20Hz is technical,
and already discussed in Sec.~\ref{sec:memoryA}. The second truncation
arises because the available waveforms include only a finite number of
modes. Therefore, instead of the full summation in
Eq.~(\ref{eq:hmode}), we have a partial expression
\begin{equation}
  \label{eq:hmode-model}
  \mathfrak{h} = \frac{1}{D_L}\,\,{\sum}^\prime_{(\ell, m)} 
  \mathfrak{h}_{\ell m}(t;\vec{\lambda})\,\, {}_{-2}Y_{\ell m}(\iota,\varphi)\,.
\end{equation}
Here the summation symbol $\Sigma^\prime$ refers to a sum only
over the available modes for the waveform model.  Thus, if
$\{(\ell_1,m_1),(\ell_2,m_2)\ldots \}$ is the list of available modes,
then
\begin{equation}
  {\sum}^\prime_{(\ell, m)}  := \sum_{\{(\ell, m)\in \{(\ell_1,m_1),(\ell_2,m_2)\ldots \}} \,.
\end{equation}
The expression for $|\dot{\mathfrak{h}}|^2$ is then similarly modified
to be a sum only over a subset of the modes obtained by combinations
of the available modes.  The expressions for the mode coefficients
$\alpha_{\ell m}$ of course remain unchanged.

In practice, the list of available modes differs for different
models.  For the physically correct waveform implied in
Eq.~(\ref{eq:hmode}) the constraint equation is, by definition, always
satisfied.  This is however not the case for the model waveform of
Eq.~(\ref{eq:hmode-model}).  Here, the constraint will generally not
be exactly satisfied and generally additional modes need to be
included in order to do so.  Moreover, the memory predicted by this
procedure will differ for different waveform approximants.  If the
predicted memory for any two appxoximants turns out be significantly
different, then it is clear that either one or both approximants can
be improved.

Once the additional modes have been included, it is of course possible
to perform a further iteration and obtain further modifications to the
waveform model.  In principle, we should continue this iteration till
we converge to a waveform which exactly satisfies the constraint.  In
practice, going beyond the first iteration is unnecessary for
gravitational wave observations (see e.g. \cite{Talbot:2018sgr} where
the second iteration is referred to as ``the memory of the memory'').
We shall restrict ourselves to the first iteration in this paper.

This has two immediate applications as discussed previously.  First,
for any gravitational wave observation, one can treat the memory as
just another inferred observable on the same footing as the final
mass, spin or the recoil velocity.  Thus, just as one obtains
posterior distributions for the masses and spins, we can calculate the
posterior distributions for the memory modes.  The second application,
independent in principle of any detections, is to use this as a
diagnostic and compare different waveform models.  

In comparing two waveform approximants following the above procedure,
there are two possible approaches. The first is to simply compute the
various memory modes $\Delta\mathfrak{h}_{\ell m}$, in a discretely
sampled parameter space. In this case, we can calculate the
differences between the memory for two approximants as a function over
parameter space. This comparison would then be a property only of the
waveform independent of any features of gravitational wave detectors
or detections (though of course, one could evaluate whether the
resulting differences could be directly measurable by any detectors).
This procedure, while straightforward and necessary, will be left to
future work.  Below we shall present the results of a different
procedure which relies on the observed merger events.  Associated to
each merger event and for each waveform approximant, it is possible to
calculate the posterior probability distributions of the model
waveform parameters.  From these posterior distributions, we can
straightforwardly calculate the posterior distribution of the
different memory modes $\Delta\mathfrak{h}_{\ell m}$.  Any differences
between these posterior distributions would indicate differences
between the underlying waveform models. Furthermore, these differences
are in a region of parameter space that is, by construction,
astrophysically relevant.  In this procedure, the original posterior
distributions of the waveform parameters depends on the gravitational
wave detectors.  The more sensitive a network (either in terms of the
detector noise properties or the network configuration) would
generally imply narrower posterior distributions.

\subsection{Results for the GWTC-1 events} 
\label{sec:memoryC}

\begin{figure*}
  \centering    
  \includegraphics[width=0.9\linewidth]{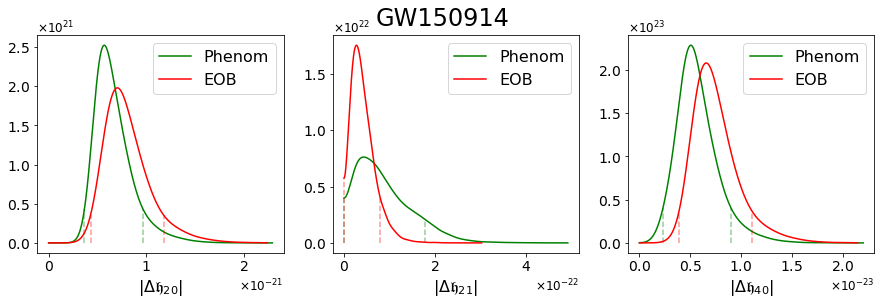}\\
  \includegraphics[width=0.9\linewidth]{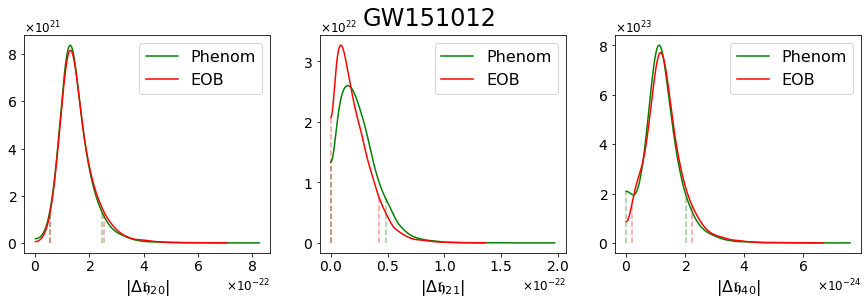}\\
  \includegraphics[width=0.9\linewidth]{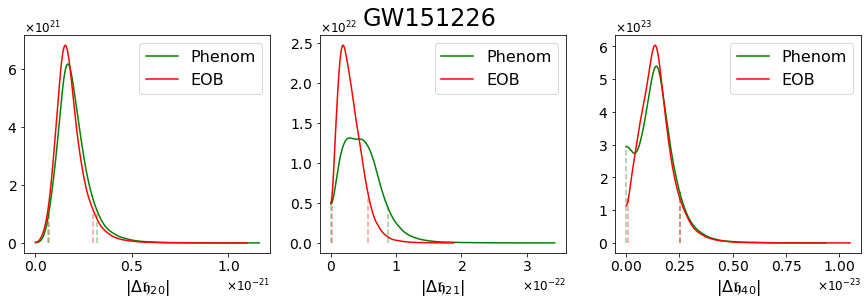}
  \caption{The posterior distributions for the total memory in the
    $(2,0),\,(2,1)$ and $(4,0)$ modes for the three O1 events.  The
    best fit values of the individual black hole masses for GW150914,
    GW151012 and GW151226 are respectively
    $(35.6 M_\odot,30.6M_\odot)$ $(13.6M_\odot, 15.2M_\odot)$ and
    $(7.7M_\odot, 8.9M_\odot)$.  As in the main text, we emphasize
    again that this not a direct measurement of the memory, but these
    are instead histograms of the inferred values of the memory
    relying on waveform models and standard general relativity.  The
    first two events are consistent with non-spinning initial black
    holes while there is some evidence for moderate spins for
    GW151226. The red and green vertical dashed lines show the $90\%$
    credible intervals (centered around the median) for the
    corresponding distribution .} \label{fig:o1events}
\end{figure*}
\begin{figure*}
  \centering    
  \includegraphics[width=0.75\linewidth]{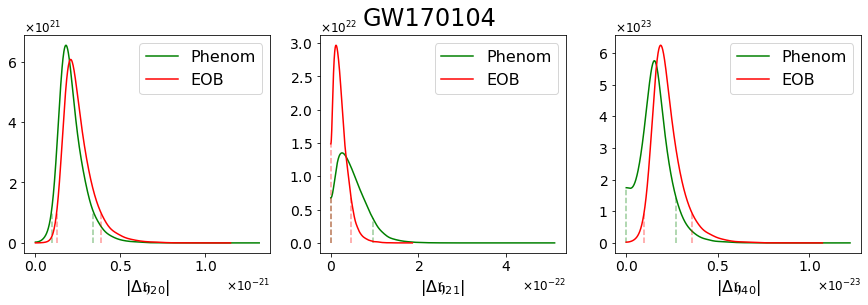}\\
  \includegraphics[width=0.75\linewidth]{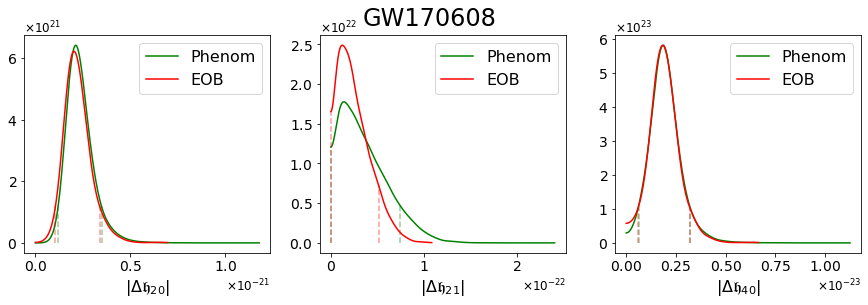}\\
  \includegraphics[width=0.75\linewidth]{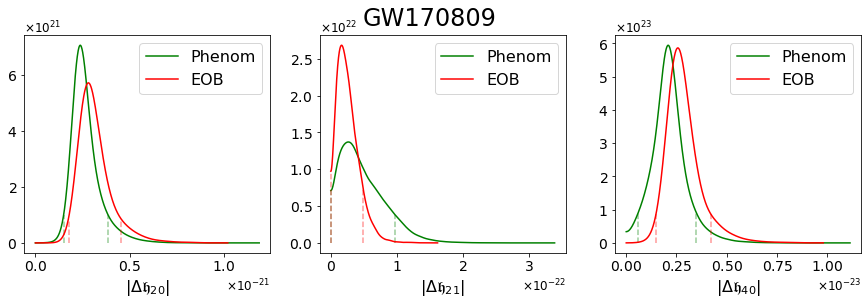}\\
  \includegraphics[width=0.75\linewidth]{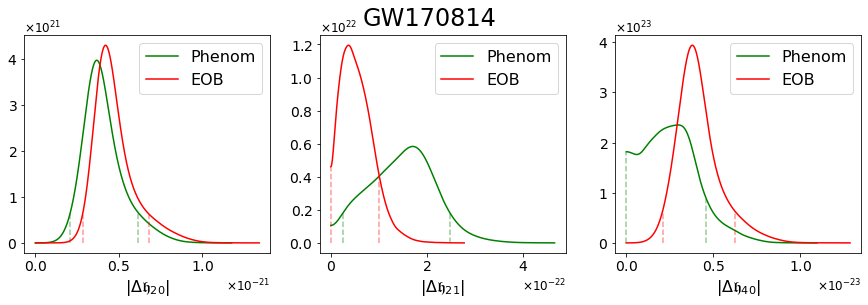}\\
  \includegraphics[width=0.75\linewidth]{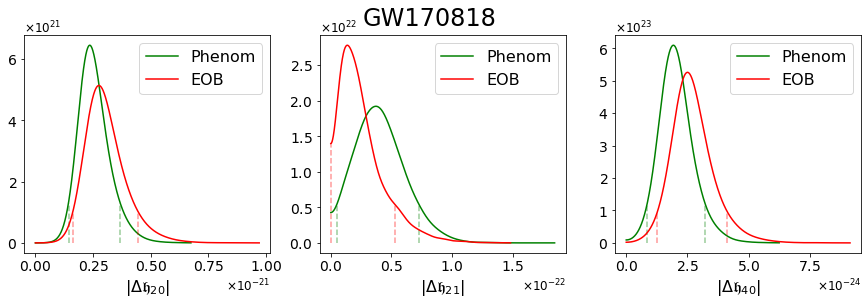}
  \caption{Posterior distributions of the inferred values of the
    memory for some selected modes. This figure presents results for
    five of the O2 events, namely GW170104, GW170608, GW170809,
    GW170814 and GW170818, in each case for both the EOB and Phenom
    models. Each row of figures corresponds to a particular event,
    while the first, second and third columns refer to the
    $(2,0), (2,1)$ and $(4,0)$ modes respectively.  These are all
    systems with moderate mass ratios, with the best fit individual
    masses being respectively $(20.0M_\odot, 21.4M_\odot)$,
    $(11.0M_\odot, 7.6M_\odot)$, $(35.0M_\odot, 23.8M_\odot)$,
    $(30.6M_\odot, 25.2M_\odot)$ and $(35.4M_\odot, 26.7M_\odot)$. All
    of these are consistent with small individual spins.  Note that
    while the posteriors of EOB and Phenom generally agree, there is a
    large difference for the $(2,1)$ mode for GW170814.  }
  \label{fig:o2-first}
\end{figure*}
\begin{figure*}
  \centering    
  \includegraphics[width=0.8\linewidth]{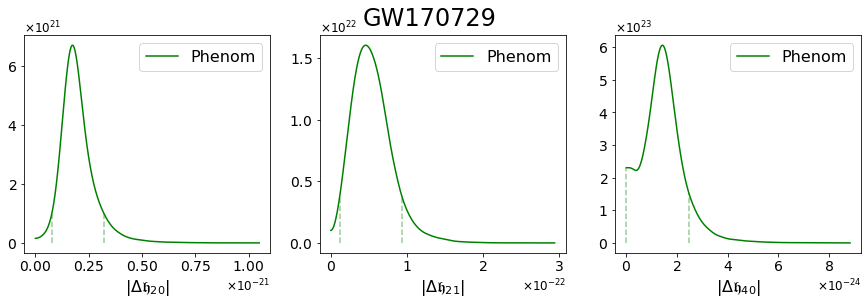}\\
  \includegraphics[width=0.8\linewidth]{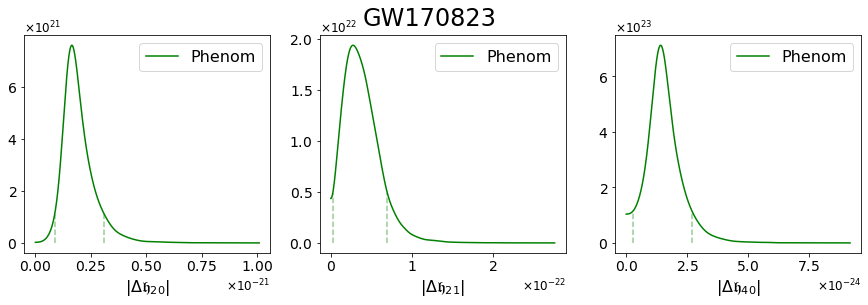}
  \caption{Posterior distributions of the inferred memory for the
    $(2,0),\,(2,1)$ and $(4,0)$ modes.  The respective modes are shown
    in the first, second and third columns for two of the O2 events
    GW170729 and GW170823, and for the IMRPhenomPv2 model. The
    individual masses for these events are respectively
    $(50.2M_\odot, 34.0M_\odot)$ and $(39.5M_\odot, 29.0M_\odot)$.
    GW170729 has moderately strong evidence of non-negligible spins. }
  \label{fig:o2-second}
\end{figure*}

We now implement this procedure for the binary mergers listed in the
first Gravitational-Wave Transient Catalog (GWTC-1)
\cite{LIGOScientific:2018mvr,Abbott:2019ebz}.  The catalog lists
binary merger events from the first and second observational science
runs of the LIGO and Virgo observatories as reported by the LIGO and
Virgo Collaborations.  The first observational run (01) covers the
duration from September 12, 2015 -- January 19, 2016 and three binary
black hole mergers are reported in GWTC-1 for this period.  The second
observational run (O2) covers the duration from November 30, 2016 --
August 25, 2017.  This period has seven binary black hole mergers and
a binary neutron star merger as reported in GWTC-1. For each of the
events, the LIGO-Virgo Collaboration has released the results of the
parameter inference studies with different waveform models, in
particular with different variants of the Phenom and EOB models.
Several other credible events apart from these have been reported in
the literature.  We note here in particular the events reported in
\cite{Venumadhav:2019tad,Zackay:2019btq,Zackay:2019tzo} and in the two
Open Gravitational Wave Catalogs (denoted 1-OGC and 2-OGC)
\cite{Nitz:2018imz,Nitz:2019hdf}.  The analysis of this paper could,
of course, be carried out for any of these additional events as well.
Since our goal is to compare different waveform models, here we use
the results from GWTC-1 only because it reports posterior
distributions for both the Phenom and EOB waveform models.
Specifically, the posterior distributions use the IMRPhenomPv2
\cite{Husa:2015iqa,Khan:2015jqa,Hannam:2013oca} and SEOBNRv3
\cite{PhysRevD.89.084006,PhysRevD.89.061502,PhysRevD.95.024010}. Both
of these are complete inspiral-merger-ringdown models including
precession.  IMRPhenomPv2 uses a single effective spin parameter while
SEOBNRv3 uses individual spins for the two black holes.  Both of these
models use only the $\ell=2$ modes, and are in fact based on applying
suitable time dependent rotations to the $\ell=2, m=\pm 2$ modes of an
underlying non-precessing model. These rotations can lead, in
principle, to all values of $m$, i.e. $-2\leq m \leq 2$ (though the
$(2,0)$ mode is generally not well modeled by a single effective spin
parameter \cite{Schmidt:2012rh}).  Thus, following the rules of
addition of angular momentum, it is easy to verify that
$|\dot{\mathfrak{h}}|^2$ (which leads to the dominant memory
contributions) contains modes with $0\leq \ell \leq 4$ and potentially
all values of $m$.

We begin with the three O1 events labeled GW150914, GW151012, and
GW151226.  For each of these events, the distributions of
$\Delta\mathfrak{h}_{20}$, $\Delta\mathfrak{h}_{21}$ and
$\Delta\mathfrak{h}_{40}$ are shown in Figs.~\ref{fig:o1events}. For
GW150914 and GW151226 there is moderate disagreement between the
Phenom and EOB results, especially for the $(2,1)$ mode.  On the other
hand, GW151012 shows excellent agreement between the different models.
The results for five of the O2 events are shown in
Fig.~\ref{fig:o2-first}.  While there are some minor discrepancies for
the $(2,0)$ and $(4,0)$ modes, it is evident that again, as for
GW150914 and GW151226, the $(2,1)$ mode shows the largest
discrepancies for several of the O2 events.  All of these five events
have moderate mass ratios and are consistent with the initial black
holes being non-spinning, and thus the sub-dominant modes due to
precession are not likely to have large amplitudes.

It is straightforward to trace back which modes of the waveform
$\mathfrak{h}_{\ell m}$ have non-vanishing contributions to a given
memory mode $\Delta\mathfrak{h}_{\ell m}$.  From
Eq.~(\ref{eq:alphalm}), we see that $\mathfrak{h}_{\ell_1 m_1}$ and
$\mathfrak{h}_{\ell_2 m_2}$ can contribute to
$\Delta\mathfrak{h}_{\ell m}$ only if
\begin{equation}
  \begin{pmatrix}  
    \ell_1 & \ell_2 & \ell \\
    m_1 & -m_2 & m
  \end{pmatrix} \neq 0\,.
\end{equation}
From the properties of the 3j-symbols we must then have $m=m_2-m_1$.
The $(2,1)$ memory mode must arise from mode combinations where $m_1$
and $m_2$ differ by unity.  An example of an allowed mode pair would
then be products of the $(2,2)$ and $(2,1)$ modes.  On the other hand,
for the $(2,0)$ or $(4,0)$ mode, we would have $m_2 = -m_1$. This
includes for example products of the $(2,2)$ and $(2,-2)$ modes which
are better modeled, unlike the $(2,1)$ mode which is generated by the
time dependent rotations mentioned above.  It is then not surprising
that $\Delta\mathfrak{h}_{2,1}$ shows the most discrepancy (However,
this argument is not entirely foolproof because the $(2,\pm 1)$ modes
contribute to $\Delta \mathfrak{h}_{2 0}$ as well, though these are
presumably generally sub-dominant).  This line of reasoning points
towards at least a general direction to resolve these differences.

Shown in Fig.~\ref{fig:o2-second} are the remaining binary black hole
merger events from O2, namely GW170729 and GW170832.  For both of
these we show only the IMRPhenomPv2 result because of the technical
difficulty related to the spin definitions at a reference frequency of
$20\,$Hz mentioned earlier.  We shall address this elsewhere and here
only present the Phenom results.


\section{Discussion}
\label{sec:discussion}

Given the required input parameters (see Section \ref{sec:intro}), EOB
and Phenom models provide us with a waveform that the detector would
receive. Therefore, in any LIGO-Virgo event, the measurement of strain
provides us with posterior probability distributions for these
parameters.  Using the commonly used terminology, we referred to them
as \emph{measured values}. Once we have these posteriors, within any
one theory, we can calculate the predicted probability distributions
for other observables in that theory.  In particular, using general
relativity, one can calculate values of the masses, spins and the
recoil velocities of the remnants. We referred to these as
\emph{inferred values}.  Future measurements with more sensitive
detectors will be able to directly measure the memory.  Once this
happens, comparison of these direct measurements with the inferred
values will yield tests of non-linear aspects of general
relativity. As shown in previous studies, this could require us to
combine $\mathcal{O}(2000)$ events \cite{Hubner:2019sly}, or wait for
the space based LISA detector \cite{Favata:2009ii}.

Before these direct measurements become reality, apart from
improvements in detector sensitivity, it will also be necessary to
improve the accuracy of waveform models.  We have shown that
differences between predictions for inferred observables made by
different waveform models can serve as indicators of differences in
the underlying physics.  In Sec.~\ref{sec:memoryA} we presented an
example to illustrate this tool: GW150914.  Although the expectation
value of the inferred observable $M_{f}$ -- the remnant mass -- in
each model is within 68\% confidence level of that in the other, the
posterior distribution in Phenom resembles a Gaussian, while that in
EOB has a double peak. We found that this difference is most likely
because of the difference in the way precession is handled in the two
models.  For the recoil velocity, both models give
inferred values that are orders of magnitude lower than
those provided by surrogate fits to numerical relativity.
This difference can be traced back directly to the fact
that, in the co-precessing frame, neither model includes
the modes that contribute to the kick. While this result
is not surprising, it provides a proof of principle that the
balance laws can be used to test accuracy of candidate
waveforms.

More importantly, the infinite tower of constraints pro- vided by the
balance laws Eq.~(\ref{eq:constraint}) can be used to compare and
contrast model waveforms.  In Sections \ref{sec:memoryB} and
\ref{sec:memoryC} we used these constraints to infer the posterior
distributions for several leading modes in the spherical harmonic
decomposition of total gravitational memory.  Memory is not an
observable of direct astrophysical interest. However, from a
fundamental general relativistic point of view, it as a gauge
invariant observable associated with the waveform. Therefore, each
spherical harmonic component of memory provides us with a new tool to
test the accuracy of waveform models. As tools, they are on the same
footing as other inferred observables such as the remnant mass and
spin.  Furthermore, these new tools can reveal discrepancies between
waveform models that were not detected by the more commonly used
observables that refer only to the properties of the remnant.
Finally, note that this analysis is rather different from discussions
of gravitational memory in the literature
\cite{Hubner:2019sly,Boersma:2020gxx} where the emphasis is on a
\emph{definitive or direct measurement} of gravitational memory from
combining several detections.  We do not address this interesting
issue.  By contrast, as we have emphasized, our goal is to regard
memory as an \emph{inferred observable} and use it to probe systematic
errors between different waveform models. In particular, our analysis
makes a strong use of general relativity because our focus is on
testing the accuracy of the candidate waveforms vis-a-vis predictions
of exact general relativity.

There are, however, some limitations of this procedure.  Once we
obtain an event for which the posterior distributions between the
models show clear systematic difference, we know that both models
cannot be good approximations to exact general relativity in a
certain region of the parameter space. As the detector sensitivity
increases such pointers could serve as powerful guidelines, calling
for further examination of the physics captured by the
models. However, it is not straightforward to identify what aspects of
the waveforms are causing this difference. Thus, the evaluative role
is ‘passive’ in the sense that the pointers by themselves do not
provide clear-cut directions to improve the models.

Nevertheless, some preliminary conjectures can be made.  The first is,
as noted in Sec.~\ref{sec:memoryC}, the flux contains products of two
modes. Therefore, given pairs $\mathfrak{h}_{\ell_1 m_1}$ and
$\mathfrak{h}_{\ell_2 m_2}$, we can identify which mode pair contributes
to each $\Delta \mathfrak{h}_{\ell m}$.  The second comment is that even
if the dominant modes (typically $(2,\pm 2)$) are well modeled, there
is still a non-trivial issue, namely, that of correlations between
various modes.  These are necessary to calculate
$|\dot{\mathfrak{h}}|^2$ accurately according to Eq.~(\ref{eq:hdotsq})
and thus greatly impact the memory.  Clearly, larger the precession or
more asymmetric the system, larger larger will the impact of the other
modes be.  It is likely that these configurations will also generally
have larger disagreements in the inferred values of the memory in
different models.  Note that both the EOB and Phenom models do not
directly model precession.  They both start with an underlying
non-precessing model to which the precession effects are applied as
suitable time dependent rotations
\cite{Schmidt:2010it,Boyle:2011gg,Schmidt:2012rh} (see also
\cite{Hannam:2013pra}).  It is generally only the underlying
non-precessing models which are directly calibrated with numerical
relativity waveforms, and the other modes are generated by the time
dependent rotations.  Thus, it is possible that if precession effects
and the higher modes were to be directly calibrated with numerical
relativity results, the disagreements with the memory reported here
would reduced.  It is worth noting that more recent precessing Phenom
models for the higher modes labeled IMRPhenomPv3HM
\cite{London:2017bcn,Khan:2019kot} already represents an improvement
in this direction.  In this model, the $(2,1)$ mode for instance is
non-vanishing even in the co-precessing frame and is thus not
determined entirely by the time dependent rotations.  Also noteworthy
are the developments on the EOB side regarding higher models,
e.g. higher modes for non-precessing systems have been modeled in
\cite{Cotesta:2018fcv} which could form the basis for including
precession effects.  Some of the more recent events reported by the
LIGO-Virgo collaboration employs these models, and it will be
interesting to repeat the analysis of this paper for those events.

Finally, there are also systematic errors involved in our analysis of
the ten LIGO-Virgo events. The evaluation of the inferred memory for a
waveform model involves calculating the mode decomposition of the
integral $\int_{-\infty}^{\infty}|\dot{\mathfrak{h}}|^2\,dt$.
However, for events considered in this paper, the waveform models used
in the publicly available analyses, SEOBNRv3 and IMRPhenomPv2, only
include the $(2,2)$ mode in the coprecessing frame. While the physics
that is ignored may be unimportant for parameter estimation, it may be
very important for the inference of certain modes of the
memory. Therefore the inferred memory we calculate may suffer from
significant systematic errors. For example the $(2,0)$ mode of the
memory is typically ~10\% larger if higher modes are included in the
waveform, and a proper estimation of the recoil velocity requires at
least the $(2,1)$ mode. However for the \emph{comparison} of two
models that are attempting to include the same physics, these errors
should be identical and a comparison of the posterior distributions is
still meaningful.  With these limitations in mind, we find that the
$(2,0)$ mode and $(2,1)$ mode are most significant sources of
memory. A comparison of these modes for the GWTC-1 events we analyzed
show that they largely agree across models - indicating that the
systematics are mostly under control. However for GW170814 we see that
the inferred $(2, 1)$ mode of the memory differs significantly between
EOB and Phenom models.

There are several interesting avenues to take this work forward.
First, there are also angular momentum balance laws
\cite{Ashtekar:2019rpv} analogous to those for supermomentum, used in
this paper. Following a procedure analogous to that of Section
\ref{sec:memoryA}, they can be used to construct posterior probability
distributions for the spin of the final remnant for any given
waveform. For any one waveform, a comparison of this distribution with
that provided by numerical relativity provides another measure of the
accuracy of that waveform. Similarly, comparisons between the
posterior distributions from two different waveform models can serve
as additional and distinct tests of the differences between the
physical underpinning. Returning to gravitational memory, an obvious
avenue is to extend our analysis to the more recently reported merger
events from the third LIGO-Virgo observational run which includes
events with higher masses and more asymmetric mass ratios
\cite{Abbott:2020tfl,LIGOScientific:2020stg}. These events could allow
for more stringent comparisons between the most up-to-date waveform
models.  As mentioned above, apart from looking at particular events,
it would be useful to compare waveform models using the memory across
large parameter space regions.  Injections of gravitational waves
spanning the parameter space can be performed to learn where the
systematic differences are prominent. Alternatively one could avoid
parameter estimation results altogether and directly compare
deviations in the inferred memory across parameter space.
Additionally once we identify the regions of parameter space where
these differences arise, such as what we see for GW170814, one can
take sample points from the region and directly compare the waveforms
and all the components that go into the inference. This should allow
one to pinpoint more accurately the source of the deviations between
the models, giving more direct input to improve future modeling.

Finally, in this paper, we focused just on total memory that involves
integrals from $t = -\infty$ to $t = \infty$ (see
Eq.~(\ref{eq:ellm})). As pointed out in \cite{Ashtekar:2019viz}, there
are also finite time versions of balance laws that enable one to
calculate the memory as a function of time, not just the difference
between very late and very early times.  This involves an accurate
calculation of, say, the $(2,0)$ and $(4,0)$ modes and a better
understanding of $\Psi_2$ \cite{Mitman:2020pbt}.  These calculations
will lead to much more detailed accuracy tests on waveforms.  Longer
term, over the next decade, as more sensitive detectors are
commissioned and more accurate waveform models are developed and the
memory is observed directly, the most important payoff will be to
compare the inferred values of the memory modes with the observed
values, thereby providing a test of non-linear general relativity.


\begin{acknowledgments}
  
  This work was supported by the NSF grants PHY-1505411 and
  PHY-1806356 and the Eberly Chair funds of Penn State. We thank
  Alessandra Buonanno, Frank Ohme and Eric Thrane for discussions and
  comments.

  We acknowledge the use of the LAL Simulation \cite{lalsuite} and
  PyCBC \cite{alex_nitz_2020_3993665} Software Packages in this paper.
 
  This research has made use of data, software and/or web tools
  obtained from the Gravitational Wave Open Science Center
  (https://www.gw-openscience.org), a service of LIGO Laboratory, the
  LIGO Scientific Collaboration and the Virgo Collaboration. LIGO is
  funded by the U.S. National Science Foundation. Virgo is funded by
  the French Centre National de Recherche Scientifique (CNRS), the
  Italian Istituto Nazionale della Fisica Nucleare (INFN) and the
  Dutch Nikhef, with contributions by Polish and Hungarian institutes.

\end{acknowledgments}

\bibliography{memory}{}
\end{document}